\documentstyle[12pt,epsfig]{article}
\begin{document}

\begin{flushright}
{\bf hep-ph/0102295} \\
{\bf LMU-01-02} 
\end{flushright}

\vspace{0.5cm}

\begin{center}
{\large\bf Bounds on $\sin 2\beta$ and $|V_{ub}/V_{cb}|$ from 
the Light-Quark Triangle}
\end{center}

\vspace{0.1cm}

\begin{center}
{\bf Harald Fritzsch} ~ and ~ {\bf Zhi-zhong Xing
\footnote{Electronic address: xing@theorie.physik.uni-muenchen.de}} \\ 
{\it Sektion Physik, Universit$\it\ddot{a}$t M$\it\ddot{u}$nchen,
Theresienstrasse 37A, 80333 M$\it\ddot{u}$nchen, Germany} 
\end{center}

\vspace{2.5cm}

\begin{abstract}
With the help of the light-quark triangle, which is essentially congruent 
to the rescaled unitarity triangle for a variety of textures of the quark 
mass matrices, we calculate the CP-violating quantity $\sin 2\beta$ and
the ratio of $|V_{ub}|$ to $|V_{cb}|$. We find that $\sin 2\beta$ is most 
likely to lie in the range $0.45 \leq \sin 2\beta \leq 0.60$, a result
compatible very well with the present BaBar and Belle measurements. 
On the other hand, $|V_{ub}/V_{cb}| \geq 0.8$ is disfavored. Our bounds on 
both $\sin 2\beta$ and $|V_{ub}/V_{cb}|$ can soon be confronted with more 
precise data to be accumulated from the asymmetric $B$-meson factories.
\end{abstract}


\newpage

Recently the BaBar and Belle Collaborations have updated their measurements of 
$\sin 2\beta$, where 
$\beta \equiv \arg [-(V^*_{cb}V_{cd})/(V^*_{tb}V_{td})]$ is an inner
angle of the unitarity triangle of quark flavor mixing \cite{PDG}, 
from the CP-violating asymmetry in 
$B^0_d$ vs $\bar{B}^0_d \rightarrow J/\psi K_S$ decays:
\begin{equation}
\sin 2\beta \; = \; \left \{ \matrix{
0.34 \pm 0.20 ({\rm stat}) \pm 0.05 ({\rm syst}) \; , 
~~~ ({\rm BaBar} ~ \cite{BaBar}) \; , \cr\cr
0.58^{+0.32}_{-0.34} ({\rm stat})^{+0.09}_{-0.10}({\rm syst}) \; , 
~~~ ({\rm Belle} ~ \cite{Belle}) \; . ~~~~~~~~~}
\right .
\end{equation}
These results are lower than, but not in conflict with the previous result
reported by the CDF Collaboration: $\sin 2\beta = 0.79 \pm 0.42$ \cite{CDF}; 
and they are also compatible with the results obtained from global analyses of 
the unitarity triangle in the standard model \cite{Stocchi}. 
Although the central value from the 
BaBar measurement is relatively lower than those from the Belle and
CDF measurements, there is no serious discrepancy that one can
really claim. In comparison with the preliminary data announced last 
year by the BaBar and Belle Collaborations \cite{Old}, their present data 
have considerably narrowed the possible room for new physics to manifest 
itself in the CP-violating asymmetry between $B^0_d \rightarrow J/\psi K_S$ 
and $\bar{B}^0_d \rightarrow J/\psi K_S$ decays \cite{NP}.
A new window is on the other hand being opened, with the help of more
precise data on $\sin 2\beta$ and other CP-violating parameters 
to be accumulated at the $B$-meson factories, towards stringent tests of the 
texture of quark mass matrices. Reliably quantitative information on quark 
mass matrices will shed light on the underlying flavor symmetry and its
breaking mechanism, which are crucial for our deeper understanding of
the origin of quark masses, flavor mixing, and CP violation.

The main purpose of this paper is to determine
$\sin 2\beta$ from the so-called light-quark triangle, whose shape depends only upon
the flavor mixing between $(u, c)$ and $(d, s)$ quarks in the heavy quark 
limit \cite{FX97}. As shown in Ref. \cite{FX99}, the light-quark triangle is 
essentially congruent to the rescaled unitarity triangle for a variety 
of realistic quark mass matrices. Therefore it is possible to calculate
the angles of the unitarity triangle, which are observable parameters of
CP violation, from the sides of the light-quark triangle. We find that
the numerical prediction of $\sin 2\beta$ from the light-quark triangle
is very well consistent with the present BaBar and Belle data. We also obtain
a very instructive bound on $|V_{ub}/V_{cb}|$, although it is somehow 
lower than the currently most favorable experimental value. Our results
of $\sin 2\beta$ and $|V_{ub}/V_{cb}|$ can soon be confronted with more 
precise data to be accumulated at the KEK and SLAC $B$-meson factories.

Let us start with a brief retrospection of the light-quark triangle
derived from the quark mass matrices with specific texture zeros.
In the standard model or its extensions which have no flavor-changing 
right-handed currents, one can always choose a specific flavor basis in which
both the up-type quark mass matrix $M_{\rm u}$ and its down-type counterpart
$M_{\rm d}$ are Hermitian and have vanishing
(1,3) and (3,1) elements \cite{FX97}. Such a flavor basis is quite natural
in the sense that it coincides with the observed hierarchy of quark masses.
Without loss of generality, the (1,1) element of $M_{\rm u}$ or $M_{\rm d}$
can also be arranged to vanish through a proper but physically 
irrelevant transformation of the chosen flavor basis \cite{Branco}. 
It is impossible, however, to arrange the (1,1), (1,3) and 
(3,1) elements of both $M_{\rm u}$ and $M_{\rm d}$ to vanish in the most 
general case. Hence quark mass matrices of the form
\begin{equation}
M_{\rm q} \; = \; \left ( \matrix{
0	& D_{\rm q}	& 0 \cr
D^*_{\rm q}	& C_{\rm q}	& B_{\rm q} \cr
0	& B^*_{\rm q}	& A_{\rm q} \cr} \right ) \; ,
\end{equation}
where q = u (up-type) or d (down-type), keep the essential generality 
except for assuming the simultaneous vanishing of the (1,1)
elements in $M_{\rm u}$ and $M_{\rm d}$. In view of the strong mass
hierarchy in each quark sector, one naturally expects that 
$|A_{\rm q}|$ is dominant over $|B_{\rm q}|$, $|C_{\rm q}|$ and $|D_{\rm q}|$ 
in magnitude. It turns out
that the heavy quark limit (i.e., $m_t \rightarrow \infty$ and 
$m_b\rightarrow \infty$), which allows the light quarks $(u,c)$ or $(d,s)$
to be decoupled from the massive $t$ or $b$ quark, is a useful and realistic 
approximation. Then the flavor mixing matrix element $|V_{us}|$ or 
$|V_{cd}|$ can be derived from the mismatch between the diagonalization of 
$M_{\rm u}$ and that of $M_{\rm d}$:
\begin{equation}
|V_{cd}| \; = \;
\left | R_{\rm u} ~ - ~ R_{\rm d} \exp (i\omega) \right | \; ,
\end{equation}
where
\begin{eqnarray}
R_{\rm u} & = & \sqrt{\frac{m_u}{m_u + m_c}} 
\sqrt{\frac{m_s}{m_d + m_s}} \;\; ,
\nonumber \\
R_{\rm d} & = & \sqrt{\frac{m_c}{m_u + m_c}} 
\sqrt{\frac{m_d}{m_d + m_s}} \;\; ,
\end{eqnarray}
and $\omega \equiv \arg (D_{\rm d}) - \arg (D_{\rm u})$. 
Such an instructive relation was discussed long time ago \cite{FX78}
to interpret the Cabibbo mixing between $(u,c)$ and $(d,s)$ quarks. Note that
Eq. (3) defines a triangle in the complex plane, the so-called light-quark
triangle as illustrated in Fig. 1(a). This triangle has been shown to
be approximately congruent to the rescaled unitarity triangle 
in Fig. 1(b) \cite{FX97,FX99}, defined by the relation
\begin{equation}
|V_{cd}| \; =\; \left |S_{\rm u} ~ - ~ S_{\rm d} \exp (i\alpha) \right | \; ,
\end{equation}
where $S_{\rm u} = |V^*_{ub}V_{ud}/V^*_{cb}|$,
$S_{\rm d} = |V^*_{tb}V_{td}/V^*_{cb}|$, and 
$\alpha \equiv \arg [-(V^*_{tb}V_{td})/(V^*_{ub}V_{ud})]$ is another 
inner angle of the unitarity triangle. As a result, the phase parameter
$\omega$, which is only relevant to the magnitude of flavor mixing between 
$(u, c)$ and $(d, s)$ quarks in the heavy quark limit,
may lead to CP violation (i.e., $\omega \approx \alpha$) once the heavy 
quark limit is slightly lifted. One can then make use of the light-quark
triangle to calculate the angles of the unitarity triangle in a good
approximation. Since the former only involves 
$|V_{cd}|$, $m_u/m_c$ and $m_d/m_s$, it is possible to predict the value of 
$\sin 2\beta$ with rather small numerical uncertainties.  
\begin{figure}[t]
\vspace{0cm}
\epsfig{file=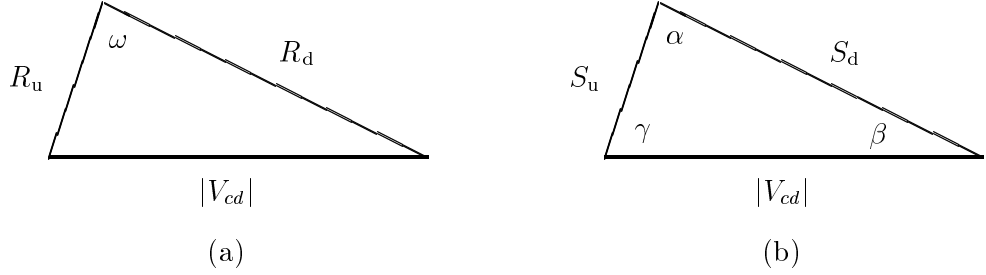,bbllx=3.3cm,bblly=6cm,bburx=19cm,bbury=28cm,%
width=15cm,height=20cm,angle=0,clip=}
\vspace{-16.1cm}
\caption{\small In the complex plane: (a) the light-quark triangle (LT); and
(b) the rescaled unitarity triangle (UT).}
\end{figure}

A particularly interesting case is $\omega = 90^\circ$; 
i.e., the light-quark triangle is a right-angled triangle \cite{FX95,Randhawa}. 
In this special case, we can estimate the magnitude of $\sin 2\beta_{\rm LT}$ 
by use of a rather simple relation:
\begin{equation}
\tan \beta_{\rm LT} \; = \; \frac{R_{\rm u}}{R_{\rm d}}
\; \approx \; \sqrt{\frac{m_um_s}{m_cm_d}} \;\; .
\end{equation}
Taking $m_u/m_c = 4\cdot 10^{-3}$ and $m_d/m_s = 0.05$ typically \cite{PDG}, 
we obtain $\beta_{\rm LT} \approx 15.8^\circ$ or 
$\sin 2\beta_{\rm LT} \approx 0.52$. The latter is fairly consistent with
experimental data given in Eq. (1). 
\begin{figure}[t]
\vspace{-0.32cm}
\epsfig{file=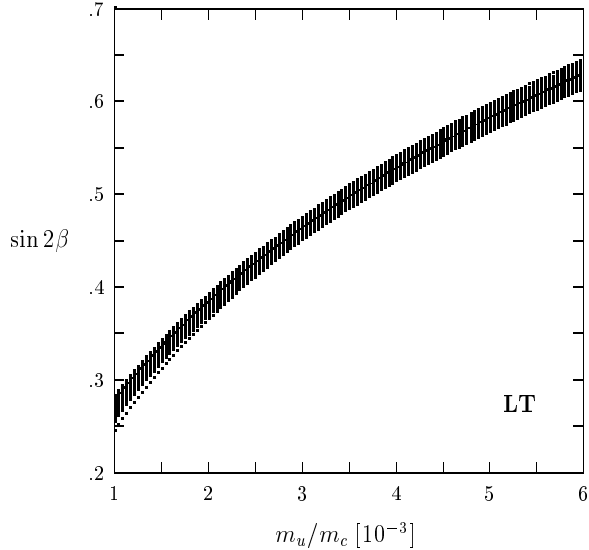,bbllx=-2.5cm,bblly=6cm,bburx=17cm,bbury=28cm,%
width=15cm,height=17.5cm,angle=0,clip=}
\vspace{-9.5cm}
\caption{\small Bound on $\sin 2\beta$ from the light-quark triangle (LT),
where $|V_{cd}| = 0.222 \pm 0.009$ and $m_s/m_d = 18.9 \pm 0.8$ 
have been input. The solid curve corresponds to the central values
of the input parameters.}
\end{figure}

More generally, one may calculate the CP-violating angle $\beta$ from the 
light-quark triangle with the help of the cosine theorem. We obtain
\begin{equation}
\cos \beta^{~}_{\rm LT} \; =\; \frac{1}{2} \sqrt{\frac{m_s}{m_d}}
\left [ |V_{cd}| + \frac{1}{|V_{cd}|} 
\left ( \frac{m_d}{m_s} - \frac{m_u}{m_c} \right ) + \Delta_{\rm LT} \right ] \; 
\end{equation}
in the next-to-leading order approximation, where
\begin{equation}
\Delta_{\rm LT} \;\; =\;\; \frac{1}{2} \left [ |V_{cd}| - \frac{1}{|V_{cd}|} 
\left ( \frac{m_d}{m_s} - \frac{m_u}{m_c} \right ) \right ]
\left ( \frac{m_u}{m_c} + \frac{m_d}{m_s} \right ) \; .
\end{equation}
Then a numerical prediction for $\sin 2\beta^{~}_{\rm LT}$ as a function of
$m_u/m_c$ can be made by taking $|V_{cd}| = 0.222 \pm 0.009$ 
\footnote{Note that we have adopted the average of the experimental values
$|V_{cd}| = 0.224 \pm 0.016$ and $|V_{us}| = 0.2196 \pm 0.0023$ \cite{PDG}
as the input of $|V_{cd}|$. The reasons are simply that 
(a) $|V_{cd}| = |V_{us}|$ holds in the heavy quark limit; and (b)
$|V_{us}| - |V_{cd}| \sim {\cal O}(10^{-4})$ holds in reality, as 
guaranteed by the unitarity of the quark flavor mixing matrix \cite{Xing95}.}
and $m_s/m_d = 18.9 \pm 0.8$ \cite{Leutwyler}. Note that the ratio $m_s/m_d$,
unlike the ratio $m_u/m_c$, can be determined rather accurately using the chiral 
perturbation theory. We plot the result in Fig. 2.
We see that the magnitude of $\sin 2\beta_{\rm LT}$ increases monotonically with
the value of $m_u/m_c$. Corresponding to the generous range of $m_u/m_c$
(i.e., $1\cdot 10^{-3} \leq m_u/m_c \leq 6\cdot 10^{-3}$), $\sin 2\beta_{\rm LT}$
takes values from 0.25 to 0.65. The uncertainties resulting from the errors
of $m_s/m_d$ and $|V_{cd}|$ are insignificant, as illustrated in Fig. 2.
It is obvious that the bound on $\sin 2\beta$ from the light-quark triangle
is very well compatible with the present BaBar and Belle data. 

Note that $\sin 2\beta \geq 0.50$ is expected to hold in the standard 
model with current data \cite{Buras}. This lower bound implies 
$m_u/m_c \geq 3.5\cdot 10^{-3}$.
Indeed the most probable values of $m_u/m_c$ lie in the
range $3\cdot 10^{-3} \leq m_u/m_c \leq 5\cdot 10^{-3}$ \cite{PDG}.
Accordingly we arrive at a rather narrow range for $\sin 2\beta$ from the
light-quark triangle: $0.45 \leq \sin 2\beta_{\rm LT} \leq 0.6$. Such an interesting 
range can be further narrowed, once our knowledge on the mass ratio
$m_u/m_c$ is improved \cite{Hall}. After more precise data are accumulated from
the $B$-meson factories at KEK and SLAC, it will be possible to
make a stringent test of $\sin 2\beta$ derived from the light-quark triangle.
\begin{figure}[t]
\vspace{-0.33cm}
\epsfig{file=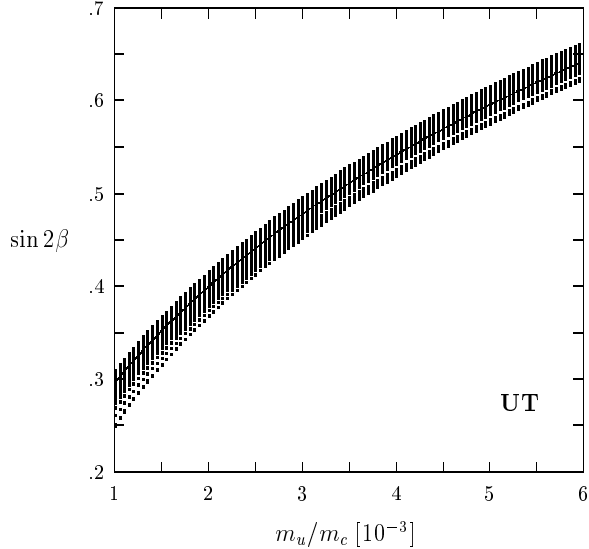,bbllx=-2.5cm,bblly=6cm,bburx=17cm,bbury=28cm,%
width=15cm,height=17.5cm,angle=0,clip=}
\vspace{-9.5cm}
\caption{\small Bound on $\sin 2\beta$ from the rescaled unitarity triangle (UT),
where $|V_{cd}| = 0.222 \pm 0.009$, $m_s/m_d = 18.9 \pm 0.8$ and
$m_b/m_s = 34 \pm 4$ have been input. The solid curve corresponds 
to the central values of the input parameters.}
\end{figure}

Now we discuss the possibility to determine $\sin 2\beta$ directly from the 
unitarity triangle, based on the quark mass matrices in Eq. (2). 
The specific deviation of the rescaled unitarity triangle 
in Fig. 1(b) from the light-quark triangle in Fig. 1(a) cannot be calculated, 
unless a further assumption is made for
the texture of $M_{\rm u}$ and $M_{\rm d}$. In a number of phenomenological
models with natural flavor symmetries \cite{FXReview}, 
$|B_{\rm q}| \sim |C_{\rm q}|$ and $|B_{\rm u}|/C_{\rm u} = |B_{\rm d}|/C_{\rm d}$
have been assumed. In this case, we obtain 
\begin{eqnarray}
S_{\rm u} & = & \sqrt{\frac{m_u}{m_c}} \left [ 1 - \frac{1}{2} |V_{cd}|^2
+ \frac{1}{2} \frac{m_c m_s}{m_u m_b} \left ( \frac{m_u}{m_c} +
\frac{m_d}{m_s} - |V_{cd}|^2 \right ) \right ] \; ,
\nonumber \\
S_{\rm d} & = & \sqrt{\frac{m_d}{m_s}} \left ( 1 + \frac{1}{2} \frac{m_u}{m_c}
- \frac{1}{2} \frac{m_d}{m_s} \right ) \; 
\end{eqnarray}
in the next-to-leading order approximation \cite{FX99}. We see that
$S_{\rm u} \approx R_{\rm u}$ and $S_{\rm d} \approx R_{\rm d}$ hold
to the leading-order degree of accuracy \cite{FX95}. Applying the
cosine theorem to Fig. 1(b) leads to 
\begin{equation}
\cos \beta^{~}_{\rm UT} \; =\; \frac{1}{2} \sqrt{\frac{m_s}{m_d}}
\left [ |V_{cd}| + \frac{1}{|V_{cd}|} 
\left ( \frac{m_d}{m_s} - \frac{m_u}{m_c} \right ) + \Delta_{\rm UT} \right ] \; 
\end{equation}
in the same order approximation, where
\begin{equation}
\Delta_{\rm UT} \; =\; \Delta_{\rm LT} ~ - ~ 
\frac{1}{2 |V_{cd}|} \sqrt{\frac{m_s}{m_d}} \left ( \frac{m_u}{m_c} + 
\frac{m_d}{m_s} - |V_{cd}|^2 \right ) \frac{m_s}{m_b} \;\; .
\end{equation}
It is clear that $\Delta_{\rm UT} = \Delta_{\rm LT}$ appears 
in the limit $m_b \rightarrow \infty$. With the inputs given above as well as
 $m_b/m_s = 34 \pm 4$ \cite{Narison},
one can similarly compute $\sin 2\beta^{~}_{\rm UT}$ as a function of
$m_u/m_c$. The numerical result is illustrated in Fig. 3.
We find that the magnitude of $\sin 2\beta_{\rm UT}$ is slightly larger than
that of $\sin 2\beta_{\rm LT}$, as a consequence of the new correction term
induced by the mass ratio $m_s/m_b$ in $\Delta_{\rm UT}$. Taking 
$m_u/m_c = 4\cdot 10^{-3}$ for example, we obtain 
$0.52 \leq \sin 2\beta_{\rm UT} \leq 0.56$ in contrast with 
$0.51 \leq \sin 2\beta_{\rm LT} \leq 0.54$. This confirms that one may
use the sides of the light-quark triangle to calculate the angles of the unitarity 
triangle to a good degree of accuracy, without involving further details of 
the quark mass matrices. 
\begin{figure}[t]
\vspace{-0.33cm}
\epsfig{file=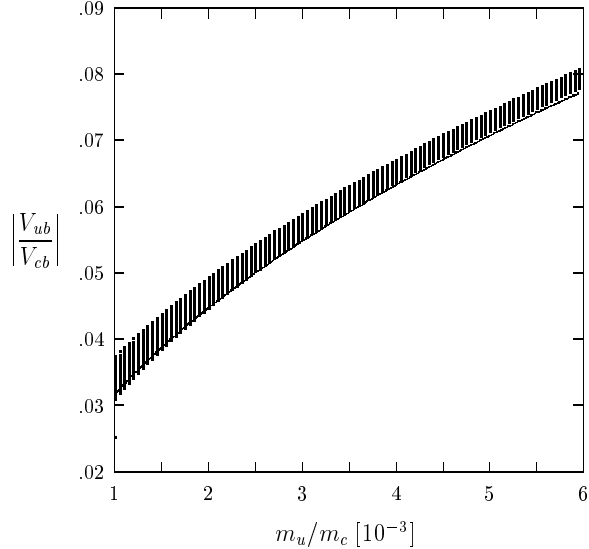,bbllx=-2.5cm,bblly=6cm,bburx=17cm,bbury=28cm,%
width=15cm,height=17.5cm,angle=0,clip=}
\vspace{-9.5cm}
\caption{\small Bound on $|V_{ub}/V_{cb}|$ from the light-quark 
triangle (the solid curve) and the rescaled unitarity triangle
(the shaded region),
where $|V_{cd}| = 0.222 \pm 0.009$, $m_s/m_d = 18.9 \pm 0.8$ and
$m_b/m_s = 34 \pm 4$ have been input.} 
\end{figure}
 
Note that an experimental value of $\sin 2\beta$ lower than that obtained
from the global fit of the unitarity triangle may have important implications
on some parameters of the standard model \cite{Buras,Nir,Ellis}. In particular,
the true value of $|V_{ub}/V_{cb}|$ might be somehow smaller than the
presently most favorable value (i.e., $|V_{ub}/V_{cb}| \approx 0.09$ \cite{PDG}).
It is therefore desirable, in the near future at $B$-meson factories, 
to check the self consistency
between the experimental measurements of $\sin 2\beta$ and $|V_{ub}/V_{cb}|$ within
the framework of the standard model. Given the texture of quark
mass matrices in Eq. (2), the magnitude of $|V_{ub}/V_{cb}|$ can be calculated
from either the light-quark triangle 
\footnote{Note that a calculation of $|V_{ub}/V_{cb}|$ from the light-quark
triangle makes sense only when the heavy quark limit is slightly lifted.
In this case, $|V_{ub}/V_{cb}|_{\rm LT}$ should be understood as the
leading-order approximation of $|V_{ub}/V_{cb}|_{\rm UT}$ determined from
the unitarity triangle.}
or the unitarity triangle. We obtain
\begin{eqnarray}
\left | \frac{V_{ub}}{V_{cb}} \right |_{\rm LT} & = & 
\sqrt{\frac{m_u}{m_c}} \;\; ,
\nonumber \\
\left | \frac{V_{ub}}{V_{cb}} \right |_{\rm UT} & = & \sqrt{\frac{m_u}{m_c}} 
\left [ 1 + \frac{1}{2} \frac{m_c m_s}{m_u m_b} \left ( \frac{m_u}{m_c}
+ \frac{m_d}{m_s} - |V_{cd}|^2 \right ) \right ] \; 
\end{eqnarray}
in the next-to-leading order approximation. Clearly $|V_{ub}/V_{cb}|_{\rm UT}$
is a little larger than $|V_{ub}/V_{cb}|_{\rm LT}$, due to the correction
from $m_s/m_b$. One may easily check, with the help of Eqs. (9) and (12), that
$|V_{ub}/V_{cb}|_{\rm UT} = S_{\rm u}/|V_{ud}|$ holds to the same degree of
accuracy. The numerical results of
$|V_{ub}/V_{cb}|_{\rm LT}$ and $|V_{ub}/V_{cb}|_{\rm UT}$ are shown in Fig. 4.
We observe that the possibility of $|V_{ub}/V_{cb}| \geq 0.08$ is 
strongly disfavored. For $m_u/m_c$ changing from $3\cdot 10^{-3}$ to
$5\cdot 10^{-3}$, we get $0.055 \leq |V_{ub}/V_{cb}|_{\rm UT} \leq 0.074$.
The experimental data obtained in the LEP experiments indicate that 
$|V_{ub}/V_{cb}|$ should be larger than $0.08$ \cite{LEP}. 
In our approach this 
can hardly be accommodated. We conclude that the LEP results for 
$|V_{ub}/V_{cb}|$ should be questioned. The issue will soon be clarified
by the new experimental data to be obtained from the $B$-meson factories.

In summary, we have calculated $\sin 2\beta$ and $|V_{ub}/V_{cb}|$ from the
light-quark triangle based on a generic texture of quark mass matrices.
The results turn out to be good approximations of those obtained directly
from the unitarity triangle. We find that the present BaBar and Belle data on
$\sin 2\beta$ can well be interpreted in the context of our model
of quark masses and CP violation. More accurate numerical predictions have to
rely on further progress in determining the mass ratio $m_u/m_c$.
A crucial experimental test of the texture of quark 
mass matrices under discussion will be the improved measurement of 
$|V_{ub}/V_{cb}|$. Our bounds on both $\sin 2\beta$ and 
$|V_{ub}/V_{cb}|$ can soon be confronted with more precise data to be 
accumulated from the asymmetric $B$-meson factories at KEK and SLAC.

\underline{Acknowledgements} ~ One of us (H.F.) is indebted to D. Schaile
for a discussion on the LEP results. Z.X. would like to thank A. Buras
for his useful comments on the lower bound of $\sin 2\beta$ in the
standard model.

\end{document}